\documentclass[12pt]{article}
\usepackage{multicol}
\usepackage{amssymb}
\usepackage{amsfonts,amssymb,amsmath,amsthm}
\usepackage{epsfig}
\usepackage{slashed}
\usepackage{graphicx}
\usepackage{subfig}
\usepackage{mathrsfs}
\usepackage{bbm}
\usepackage{cite}
\usepackage{enumerate}
\usepackage{slashbox}
\usepackage{color}
\usepackage[colorlinks,linkcolor=red,anchorcolor=red,citecolor=green]{hyperref}
\makeatletter
\@addtoreset{equation}{section}
\makeatother

\begin{document}
\renewcommand{\thefootnote}{\fnsymbol{footnote}}
\begin{titlepage}

\vspace{10mm}
\begin{center}
{\Large\bf Phase transition and entropy inequality of noncommutative black holes in a new extended phase space}
\vspace{16mm}

{{\large Yan-Gang Miao\footnote{\em E-mail: miaoyg@nankai.edu.cn}
and Zhen-Ming Xu\footnote{\em E-mail: xuzhenm@mail.nankai.edu.cn}}\\

\vspace{6mm}
{\em School of Physics, Nankai University, Tianjin 300071, China}

%\vspace{3mm}
%${}^{2}${\em State Key Laboratory of Theoretical Physics, Institute of Theoretical Physics, \\Chinese Academy of Sciences, P.O. Box 2735, Beijing 100190, China}

%\vspace{3mm}
%${}^{3}${\normalsize \em CERN, PH-TH Division, 1211 Geneva 23, Switzerland}
}

\end{center}

\vspace{10mm}
\centerline{{\bf{Abstract}}}
\vspace{6mm}
We analyze the thermodynamics of the noncommutative high-dimensional Schwarzschild-Tangherlini AdS black hole with the non-Gaussian smeared matter distribution by regarding a noncommutative parameter as an independent thermodynamic variable named as  {\em the noncommutative pressure}. In the new extended phase space that includes this noncommutative pressure and its conjugate variable,  we  reveal that the noncommutative pressure and the original thermodynamic pressure related to the negative cosmological constant make the opposite effects in the phase transition of the noncommutative black hole, i.e. the former dominates the UV regime while the latter does the IR regime, respectively. In addition, by means of the reverse isoperimetric inequality, we indicate that only the black hole with the Gaussian smeared matter distribution holds the maximum entropy for a given thermodynamic volume among the noncommutative black holes with various matter distributions.

\vskip 20pt
\noindent
{\bf PACS Number(s)}: 04.50.Gh; 04.60.Bc; 04.70.Dy % 05.70.-a, 05.70.Ce, 05.70.Fh;

\vskip 10pt
\noindent
{\bf Keywords}:
Phase transition, entropy inequality, noncommutative geometry, extended phase space, non-Gaussian smeared matter distribution

\end{titlepage}

\newpage
\renewcommand{\thefootnote}{\arabic{footnote}}
\setcounter{footnote}{0}
\setcounter{page}{2}
\pagenumbering{arabic}
%\tableofcontents
%\vspace{1cm}

\section{Introduction}
As a fascinating object, a black hole is increasingly popular in the classical and quantum gravity theories~\cite{XC,BS}. Its thermodynamics has been studied~\cite{JDB1,JMB,RW,SC}, especially in the AdS spacetime where the geometry plays a pivotal role in recent developments of theoretical physics~\cite{SH,RGC}. The AdS/CFT duality~\cite{JMM,EW} offers a powerful tool to tackle nonperturbative features of a variety
of physical systems, and the holographic principle~\cite{LSLS,RB} provides a way to study dynamics of a general gravitational system. The most typical example is the area law relating the entropy of a semi-classical black hole to its horizon area.

Moreover, the introduction of a negative cosmological constant injects new vitality into the research of thermodynamics of black holes,  i.e. the variation of the cosmological constant enters the first law of thermodynamics~\cite{BPD,CEJM,KM,DSJT,BPD2}. When the cosmological constant is interpreted as the thermodynamic pressure $P$ in the equation of state,
\begin{equation}
P=-\frac{\Lambda}{8\pi}=\frac{(n-1)(n-2)}{16 \pi l^2},\label{pres}
\end{equation}
where $n$ stands for the dimension of spaetime and $l$ represents the curvature radius of the AdS spacetime, the mass of black holes is identified with the enthalpy rather than the internal energy. As is known in some more fundamental theory, physical constants,  such as the Yukawa coupling, the gauge couplings, the Newton gravitational constant, and the cosmological constant, are not fixed but  vary as  dynamical variables, which arises from vacuum expectation values. As a result, variations of these so-called constants should be included in the first law of black hole thermodynamics, as shown in the Born-Infeld black hole~\cite{GKM,NB} and the Gauss-Bonnet black hole~\cite{RCLY,XXZ}.

Recently, the thermodynamics of a noncommutative black hole has received wide attention~\cite{PNSS,ESS,ANSS,SSN,AESS,TG,RBPN,LMWZ,NM,MY,PN}.
%The basic idea to introduce a noncommutative effect of spacetime into a black hole is to keep the left hand side of Einstein's field equations unchanged, while to replace the point-like $\delta$-function mass density by the Gaussian or non-Gaussian smeared matter distributions in the right hand side. In other words, by modifying energy-momentum tensors in terms of smeared matter distributions, the noncommutative effect is imposed upon black holes. A self-regular black hole solution with noncummutative effects but without curvature singularities is thus given~\cite{PN}.
Noncommutative black holes, i.e. noncommutative geometry inspired black holes, originate from the possibility of implementing an effective minimal length responsible for delocalization of the point-like object in general relativity. The main idea is that the noncommutativity of spacetime would be an intrinsic rather than a super-imposed property of manifold. To realize such an idea, one can  modify the energy-momentum tensor in terms of smeared matter distributions, such that this modified tensor describes a kind of anisotropic fluid rather than the perfect fluid, but keep the Einstein tensor unchanged in the field equations of  gravity~\cite{RBPN}. As a result, the metric of noncommutative black holes is the solution of the modified Einstein's equations. There are two archetypical features of such noncommutative black holes. One is that no singularity exists at the origin which can be smeared out by a specific matter distribution. The other is that it naturally contains an extreme configuration of black holes with a minimal length which originates from the horizon radius of the extreme black hole. The features eliminate the unfavorable divergency of the Hawking radiation.

In this paper, based on a new extended phase space associated with a noncommutative parameter as an intensive variable, we investigate the thermodynamics of the noncommutative high-dimensional Schwarzschild-Tangherlini AdS black hole~\cite{ST} with the non-Gaussian smeared matter distribution~\cite{Park,POS,MX}. By making a dimensional analysis of the noncommutative parameter and an analogy of it with the cosmological constant~\cite{BPD,CEJM,KM,DSJT,BPD2}, the Born-Infeld parameter~\cite{GKM,NB}, and the Gauss-Bonnet coupling constant~\cite{RCLY,XXZ} that are all dealt with as a kind of thermodynamic pressure, we regard the noncommutative parameter as an independent thermodynamic variable called {\em the noncommutative pressure}. Under this assumption, such pressure will appear in the generalized Smarr relation and its variation will be included in the first law of thermodynamics.
%In this way, we can give the exact expression of thermodynamic quantities of the noncommutative black hole. as the pressure,
In particular, we find that the noncommutative pressure and the thermodynamic pressure make opposite effects in the phase transition of the noncommutative black hole, where the former corresponds to the ultra-violet physics while the latter to the infra-red physics, respectively. Moreover, we show that the reverse isoperimetric inequality~\cite{CGKP,AKMS} remains valid for the noncommutative black hole, and also indicate that the noncommutative black hole with the Gaussian smeared matter distribution holds the maximum entropy for a given thermodynamic volume.

The organization of the present paper is as follows. In section \ref{sec2}, we regard the noncommutative parameter as a new pressure variable and discuss the thermodynamic behaviors of the noncommutative black hole in the extended phase space associated with this intensive variable and its conjugate one. In section \ref{sec3}, the maximum entropy  problem of the noncommutative black hole is  discussed in detail. Finally, we make our conclusions in section \ref{sec4}. Note that the geometric units, $\hbar=c=k_{_B}=G=1$, are adopted throughout this paper.

\section{Thermodynamics of noncommutative black holes in an extended phase space}\label{sec2}
Based on noncommutative black holes incorporating effects of quantum gravity in the high energy or short distance regime of  gravitational fields, %geometrically, by means of averaging coordinate operators on suitable coordinate coherent states, calculations of the literature
the authors of ref.~\cite{ASES11} have indicated that a smeared matter density in a noncommutative manifold is governed by a Gaussian distribution, no longer by a Dirac delta function. From a physical point of view, the noncommutative geometry is described as a fluid diffused around the origin rather than squeezed at the origin. This observation gives the result that the modified energy-momentum tensor with smeared matter distributions corresponds to an anisotropic fluid rather than a conventional isotropic fluid. Moreover, the recent studies~\cite{Park,POS,MW} show that the Gaussian smeared matter distribution is not always required, as long as the distribution has a sharp peak at the origin like a Dirac delta function,  and that the integration of the distribution function takes a finite value. Hence, a general smeared matter density was proposed~\cite{Park},\footnote{The general smeared matter density was applied in three dimensions. Here we extend it to the $n$-dimensional spacetime based on the achievement~\cite{TG} for the case of $k=0$.}
\begin{equation}
\rho(r)={\tilde A}\, r^k e^{-\left(\frac{r}{2\sqrt{\theta}}\right)^2},\label{fenbu}
\end{equation}
where $\sqrt{\theta}$ is the noncommutative parameter with the dimension of length which is associated with the effective minimal length, ${\tilde A}$ is a normalization constant\footnote{The normalization constant ${\tilde A}$ can be fixed by using the constraint: $\int_0^{\infty}\rho(r) \text{d}V_{n-1}=M$, where the parameter $M$ is the ADM mass of black holes, and $\text{d}V_{n-1}$ is an $(n-1)$-dimensional volume element. One thus has the complete form of the general smeared matter density, $\rho(r)=\frac{\Gamma\left(\frac{n-1}{2}\right)}{\Gamma\left(\frac{n+k-1}{2}\right)}\frac{Mr^k e^{-r^2/(4\theta)}}{\pi^{\frac{n-1}{2}}\left(2\sqrt{\theta}\right)^{n+k-1}}$.} and $k$ is a non-negative integer, $k=0, 1, 2, \cdots$. For instance, the Gaussian distribution corresponds to $k=0$, Rayleigh distribution to $k=1$, and Maxwell-Boltzmann distribution to $k=2$, etc.

For a static and spherically symmetric noncommutative black hole, the geometry structure of solutions has to satisfy the two conditions: (i) Radial matter distribution function $\rho=\rho(r)$, which is required by spherical symmetry, and (ii) Covariant energy conservation $\nabla_{\mu}T^{\mu\nu}=0$, which is required by the effective Einstein equations eq.~(\ref{einfield}). In addition, the third condition,  (iii) Schwarzschild-like solution $g_{00}g_{rr}=-1$, is added here for discussing the simplest case. Without it, the noncommutative geometry-inspired dirty black holes can be obtained~\cite{MV,EPES12}. Thus, one can find the modified energy-momentum tensor~\cite{RBPN},
\begin{equation}
{T^{\mu}}_{\nu}=\text{Diag}(-\rho, p_r, p_{\perp}, p_{\perp}),\label{emten}
\end{equation}
which represents an anisotropic fluid with the density $\rho$, the radial pressure $p_r=-\rho$, and the tangential pressure $p_{\perp}=-\rho-(r/2)(\text{d}\rho/\text{d}r)$, rather than the perfect fluid. We must notice that the non-vanishing radial pressure would balance the inward gravitational pull and prevent the smeared matter collapsing into a point. This effect coincides accurately with the one caused by an effective minimal length in spacetime.

Therefore, we consider the modified Einstein equations with the energy-momentum tensor of an anisotropic fluid eq.~(\ref{emten}) in the AdS background,
\begin{equation}
R_{\mu\nu}-\frac12 R g_{\mu\nu}+\Lambda g_{\mu\nu}=8\pi T_{\mu\nu},\label{einfield}
\end{equation}
and the metric with the spherical symmetry in $n$ dimensions~\cite{MX},
\begin{equation}
\text{d}s^2=-f(r)\text{d}t^2+\frac{\text{d}r^2}{f(r)}+r^2 \text{d}\,{\Omega}^{2}_{n-2}, \label{metrice}
\end{equation}
where $\text{d}\,\Omega_{n-2}^2$ is the square of line element on an $(n-2)$-dimensional unit sphere. By solving eqs.~(\ref{emten})-(\ref{metrice}) together with the smeared matter density eq.~(\ref{fenbu}), one can obtain the specific form of function $f(r)$,
\begin{equation}
f(r)=1-\frac{16\pi m(r)}{(n-2)\omega r^{n-3}}+\frac{r^2}{l^2}. \label{metric}
\end{equation}
Note that $\omega$ denotes the area of  an $(n-1)$-dimensional unit sphere, $\omega=2\pi^{\frac{n-1}{2}}/\Gamma\left(\frac{n-1}{2}\right)$, and $m(r)$ is related to the mass distribution function of black holes\footnote{The mass distribution function is calculated in ref.~\cite{MX} by utilizing its definition $m(r) \equiv \int_0^{r}\rho(r^{\prime}) \text{d}V_{n-1}$ and the smeared matter distribution eq.~(\ref{fenbu}).}
\begin{equation}
m(r)=\frac{M}{\Gamma\left(\frac{n+k-1}{2}\right)}\gamma\left(\frac{n+k-1}{2},\frac{r^2}{4\theta}\right),\label{masfenbu}
\end{equation}
where $\Gamma(x)$ is the gamma function and $\gamma(a,x)$ is the lower incomplete gamma function. %Incidentally, when taking the limit $\theta\rightarrow 0$, we can see that eq.~(\ref{masfenbu}) turns back to the commutative formulation, $m(r)\rightarrow M$.
For the non-Gaussian mass density distribution (eq.~(\ref{fenbu})), the matter mean radius reads
\begin{equation}\label{meanrad}
\bar{r}=\int_0^{\infty}r\, \frac{\rho(r)}{M} dV_{n-1}=2\sqrt{\theta}\, \frac{\Gamma(\frac{n+k}{2})}{\Gamma(\frac{n+k-1}{2})}.
\end{equation}
The larger the parameter $\theta$ is, the more diffuse the matter distribution is, while the smaller the parameter $\theta$ is, the more concentrated the matter distribution is. In fact, the noncommutativity of spacetime is a small effect superposed on the ordinary spacetime if it exists. In the limit $\theta\rightarrow 0$, the matter mean radius goes to zero and the matter mass eq.~(\ref{masfenbu}) goes to the total  mass $M$,  which implies that the matter distribution collapses into a point and the noncommutativity of spacetime disappears. Hence the noncommutative effect is mainly embodied in the region nearby the matter mean radius.

The location of black hole horizons is $r_h$, which is thought to be the largest real root of $f(r)=0$. The literature~\cite{PNSS,ESS,ANSS,SSN,AESS,TG,RBPN,LMWZ,NM,MY,PN,Park,POS,MX} has pointed out that the black holes with smeared matter  distributions possess a significant feature, i.e. the existence of an extreme configuration of black holes, in short  an {\em extreme black hole}. The temperature of such a black hole vanishes, indicating that it is in a frozen state. The thermodynamic behaviors of the noncommutative black holes have been discussed  in the ordinary phase space, see our previous work~\cite{MX} for the details. Here we only mention the thermodynamic enthalpy $H=M$ for our later use. Utilizing eqs.~(\ref{metric}) and (\ref{masfenbu}), we can get the enthalpy in terms of the horizon radius of black holes $r_h$,
\begin{equation}
M=\frac{(n-2)\omega \Gamma\left(\frac{n+k-1}{2}\right)}{16\pi \gamma\left(\frac{n+k-1}{2},\frac{r_h^2}{4\theta}\right)}\left(r_h^{n-3}+\frac{r_h^{n-1}}{l^2}\right). \label{enth}
\end{equation}

We now analyze the thermodynamics of noncommutative black holes with a new perspective.

At first, let us have a closer look at the noncommutative parameter $\theta$. It is obvious that $\sqrt{\theta}$ gives the characteristic length of noncommutative spacetimes and its dimension is $\text{Length}$.
By analogy with the cosmological constant~\cite{BPD,CEJM,KM,DSJT,BPD2} and the Gauss-Bonnet coupling constant~\cite{RCLY,XXZ} that are all dealt with as a kind of thermodynamic pressure, we regard the noncommutative parameter as a new intensive thermodynamic variable $P_{NC}$ called the noncommutative pressure,
\begin{equation}
P_{NC}=\frac{1}{4\pi\theta}, \label{ncpress}
\end{equation}
which will appear in the Smarr relation. It is known that the thermodynamic pressure $P$ (eq.~(\ref{pres})) is associated with the action of the AdS background spacetime to the thermodynamic system of black holes. Similarly, the above noncommutative pressure $P_{NC}$ (eq.~(\ref{ncpress})) can be regarded as the action of the self-gravitating droplet of anisotropic fluid~\cite{PNSS,PN} to the thermodynamic system of black holes.
With the help of eq.~(\ref{enth}) and the relevant thermodynamic relation, we have its conjugate extensive variable, i.e. noncommutative volume $V_{NC}$,
\begin{equation}
V_{NC}=\left(\frac{\partial M}{\partial P_{NC}}\right)_{S,\,P}
=-\frac{(n-2)\omega \theta \Gamma\left(\frac{n+k-1}{2}\right)}{8 \gamma\left(\frac{n+k-1}{2},\frac{r_h^2}{4\theta}\right)}\left(r_h^{n-3}+\frac{r_h^{n-1}}{l^2}\right)G\left(n,k;\frac{r_h}{2\sqrt{\theta}}\right), \label{ncvol}
\end{equation}
where the function $G(n,k;x)$ is defined by
\begin{equation}
G(n,k;x):=\frac{2 x^{n+k-1} e^{-x^2}}{\gamma\left(\frac{n+k-1}{2},
x^2\right)}. \label{tezheng}
\end{equation}
Therefore, the pair of conjugate variables ($P_{NC}$, $V_{NC}$) is used in the physical system that consists of the black hole together with the self-gravitating droplet of anisotropic fluid as a background.

Secondly, we suggest that the entropy $S$ of the noncommutative black hole is still the standard Bekenstein-Hawking entropy which equals $1/4$ of the event horizon area $A$,
\begin{equation}
S =\frac{A}{4}=\frac{\omega r_h^{n-2}}{4}. \label{entro}
\end{equation}
According to eqs.~(\ref{enth}) and (\ref{entro}), we can obtain the temperature $T_h$,
\begin{eqnarray}
T_h&=&\left(\frac{\partial M}{\partial S}\right)_{P, \,P_{NC}}\nonumber \\
&=&\frac{\Gamma\left(\frac{n+k-1}{2}\right)}
{4\pi\gamma\left(\frac{n+k-1}{2},\frac{r_h^2}{4\theta}\right)}\left\{\frac{n-3-G\left(n,k;\frac{r_h}{2\sqrt{\theta}}\right)}{r_h}+\frac{r_h\left[n-1-G\left(n,k;\frac{r_h}{2\sqrt{\theta}}\right)\right]}{l^2}\right\}. \label{temp}
\end{eqnarray}
In light of eqs.~(\ref{entro}) and (\ref{temp}) we make some comments as follows.
\begin{itemize}
  \item For the black holes with smeared matter  distributions, the extreme configuration exists and its corresponding temperature vanishes. We must guarantee a vanishing entropy for the extreme black hole in accordance with the third law of thermodynamics. The entropy of the extreme black hole is a constant for a given spacetime structure and a fixed matter distribution. Such a constant can be absorbed into a redefinition of entropy. Hence, eq.~(\ref{entro}) is acceptable as the expression of the entropy of the noncommutative black hole.
  \item As the temperature and entropy are conjugate thermodynamic variables to each other, it is reasonable to first define the entropy of the noncommutative black hole as that of the commutative black hole in form only. Then we can obtain the temperature according to eq.~(\ref{temp}). In fact, both the temperature and entropy are noncommutatively corrected because the horizon radius $r_h$ is implicitly $\theta$-dependent. We note that such a treatment coincides with that of the pair of conjugate variables ($P_{NC}$, $V_{NC}$), see, for instance, eqs.~(\ref{ncpress}) and~(\ref{ncvol}). This feature is understandable because the two pairs of conjugate variables appear in the forms of $T_h dS$ and $V_{NC}dP_{NC}$ in the first law of thermodynamics.
  \item The temperature eq.~(\ref{temp}) turns back to that of the commutative black hole~\cite{BC} under the limit $\theta\rightarrow 0$, which shows the consistency of the noncommutative generalization.
\end{itemize}

At last, the thermodynamic volume $V$ corresponding to the pressure $P$ (eq.~(\ref{pres})) takes the form,
\begin{eqnarray}
V=\left(\frac{\partial M}{\partial P}\right)_{S,\,P_{NC}}
=\frac{\Gamma\left(\frac{n+k-1}{2}\right)}
{\gamma\left(\frac{n+k-1}{2},\frac{r_h^2}{4\theta}\right)}\frac{\omega}{n-1}r_h^{n-1}. \label{volu}
\end{eqnarray}
We can see that the volume $V$, like $V_{NC}$, is also noncommutatively corrected.

In accordance with the above scenario,
the first law of thermodynamics for the noncommutative black hole in the extended phase space that contains the variation of the noncommutative pressure can be written as
\begin{equation}
dM=T_h dS+VdP+V_{NC}dP_{NC}. \label{fl}
\end{equation}
From the dimensional scaling of the variables, $[M]=L^{n-3}$, $[T_h]=L^{-1}$, $[S]=L^{n-2}$, $[P]=L^{-2}$, $[V]=L^{n-1}$, $[P_{NC}]=L^{-2}$, and $[V_{NC}]=L^{n-1}$, we infer the generalized Smarr relation,
\begin{equation}
(n-3)M=(n-2)T_h S-2VP-2V_{NC}P_{NC}, \label{smarr}
\end{equation}
which is a simple consequence of the first law eq.~(\ref{fl}) in the $n$-dimensions spacetime. It can be verified easily when we
substitute eqs.~(\ref{pres}), (\ref{enth})-(\ref{ncvol}), and (\ref{entro})-(\ref{volu}) into eq.~(\ref{smarr}).

Now we have two pairs of similar conjugate thermodynamic quantities, i.e. the thermodynamic pressure and thermodynamic volume ($P$,$V$) and the noncommutative pressure and the noncommutative volume ($P_{NC}$, $V_{NC}$). Then let us make a comparison about these two pairs of conjugate variables.
\begin{itemize}
  \item We notice that the thermodynamic volume eq.~(\ref{volu}) is positive, while the noncommutative volume eq.~(\ref{ncvol}) is negative. They appear in the work terms, $PV$ and $P_{NC}V_{NC}$, respectively,  see eq.~(\ref{smarr}), and have contributions to the enthalpy (or thermodynamic characteristic function) which can be thought of as the total energy of the black hole thermodynamic system. For the negative noncommutative volume eq.~(\ref{ncvol}), it can be interpreted as the self-gravitating droplet of anisotropic fluid doing work to the thermodynamic system by pushing against the system and reducing its volume to create the noncommutative black hole via a thermodynamic process. The similar peculiar feature, i.e. the negative volume  puzzle, has also appeared in the AdS-Taub-NUT case~\cite{CVJ}. For the thermodynamic volume eq.~(\ref{volu}), it means that a certain part of space is excised to `make a place' for forming a black hole at the cost of energy $P V$ as reported in ref.~\cite{CVJ}.
  \item The thermodynamic volume eq.~(\ref{volu}) just depends on the parameters $\theta$, while the noncommutative volume eq.~(\ref{ncvol}) depends on the parameters $\theta$ and $l$. In the limit $l\rightarrow \infty$, the thermodynamic pressure goes to zero, which implies no AdS background spacetime, while the noncommutative volume becomes
      \begin{equation*}
      V_{NC}\rightarrow -\frac{(n-2)\omega \theta \Gamma\left(\frac{n+k-1}{2}\right) r_h^{n-3}}{8 \gamma\left(\frac{n+k-1}{2},\frac{r_h^2}{4\theta}\right)} \cdot G\left(n,k;\frac{r_h}{2\sqrt{\theta}}\right),
      \end{equation*}
and plays a major role in a thermodynamic process. In the limit $\theta \rightarrow 0$, although the noncommutative pressure diverges, the work term $P_{NC} V_{NC}$ goes to zero, which implies no noncommutative effect, while the thermodynamic volume eq.~(\ref{volu}) turns back to the commutative one, $V_{\text{commutative}}=\omega r_h^{n-1}/(n-1)$. Hence such an asymptotic behavior is acceptable for this new pair of conjugate variables ($P_{NC}$, $V_{NC}$) in the first law of thermodynamics under the limit $\theta \rightarrow 0$. From the minus sign of the noncommutative volume, we predict that the two pressure variables ($P$ and $P_{NC}$) will play an opposite role in the thermodynamic process, which will be confirmed in the behaviors of the Gibbs free energy.
  \item Due to the observation mentioned above that the noncommutative effect is mainly concentrated in the vicinity of the matter mean radius eq.~(\ref{meanrad}), we find that our model can be regarded as an analogue to the evaporation of liquid droplets in which the two work terms $\sigma \text{d}\mathscr{A}$ (surface tension $\sigma$ and surface area $\mathscr{A}$) and $\mathscr{P} \text{d}\mathscr{V}$ (vapor pressure $\mathscr{P}$ and volume $\mathscr{V}$) play an opposite role in order to keep the liquid droplet in equilibrium and both of them appear in the first law of thermodynamics. In our model, the corresponding two work terms are $P_{NC}\text{d}V_{NC}$ and $P \text{d}V$ that appear in the first law of thermodynamics for the noncommutative black hole, see eq.~(\ref{fl}).
\end{itemize}

Next we consider the Gibbs free energy that is defined as the Legendre transform of the enthalpy (eq.~(\ref{enth})),
\begin{equation}
G:=M(r_h)-T_h (r_h)S(r_h). \label{dgibb}
\end{equation}
By inserting eqs.~(\ref{enth}), (\ref{entro}), and (\ref{temp}) into the above definition, we can obtain the exact expression of the Gibbs free energy,
\begin{equation}
G=\frac{\omega \Gamma\left(\frac{n+k-1}{2}\right)}
{16\pi \gamma\left(\frac{n+k-1}{2},\frac{r_h^2}{4\theta}\right)}\left\{\left[G\left(n,k;\frac{r_h}{2\sqrt{\theta}}\right)+1\right]r_h^{n-3}+\left[G\left(n,k;\frac{r_h}{2\sqrt{\theta}}\right)-1\right]\frac{r_h^{n-1}}{l^2}\right\}. \label{gibb}
\end{equation}
Incidentally, the Gibbs free energy tends to the commutative formula~\cite{RD} under the limit $\theta\rightarrow 0$,
\begin{equation}
G \rightarrow \frac{\omega}{16 \pi}\left(r_h^{n-3}-\frac{r_h^{n-1}}{l^2}\right).
\end{equation}

\begin{figure}
\begin{center}
  \begin{tabular}{cc}
    \includegraphics[width=60mm]{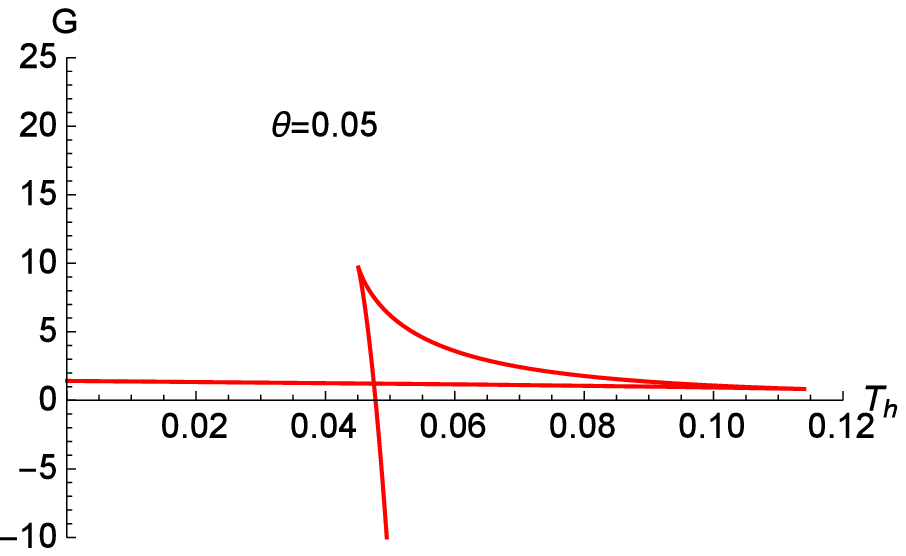} &
    \includegraphics[width=60mm]{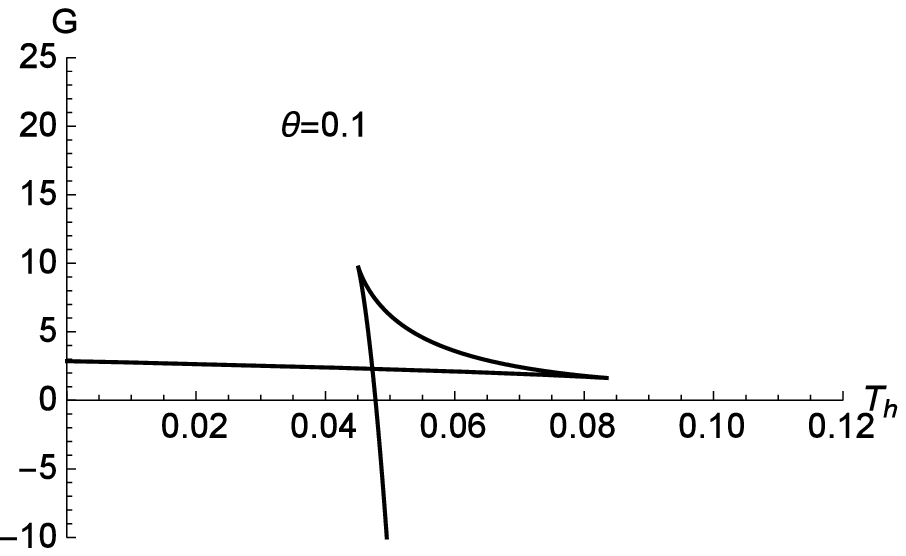} \\
    \includegraphics[width=60mm]{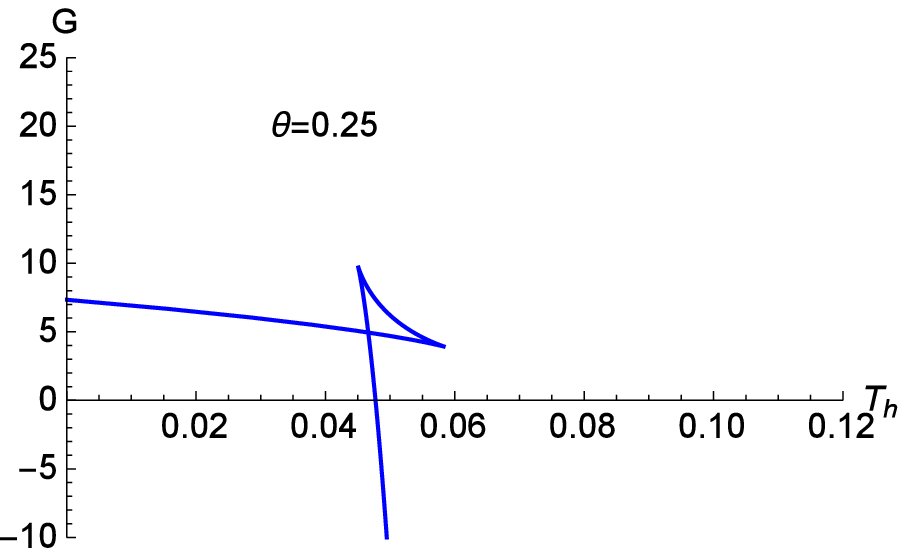} &
    \includegraphics[width=60mm]{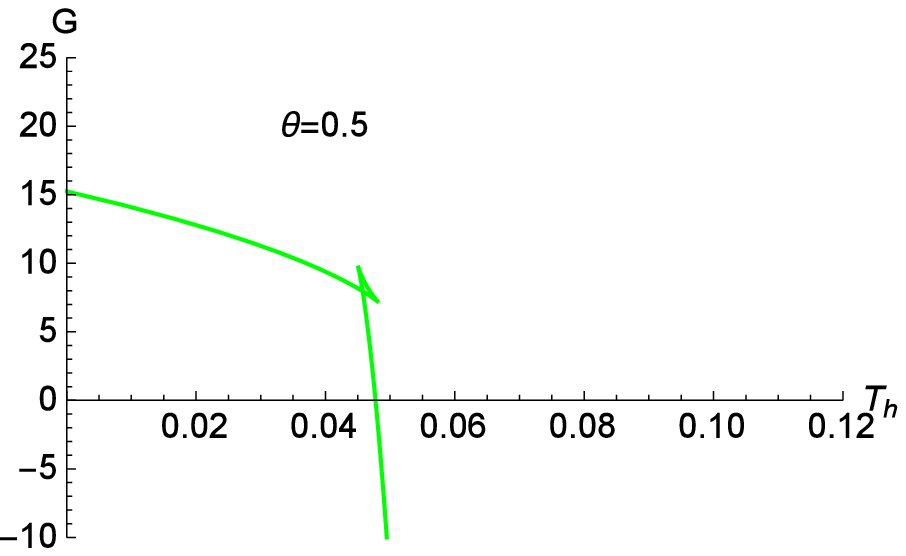} \\
    \includegraphics[width=60mm]{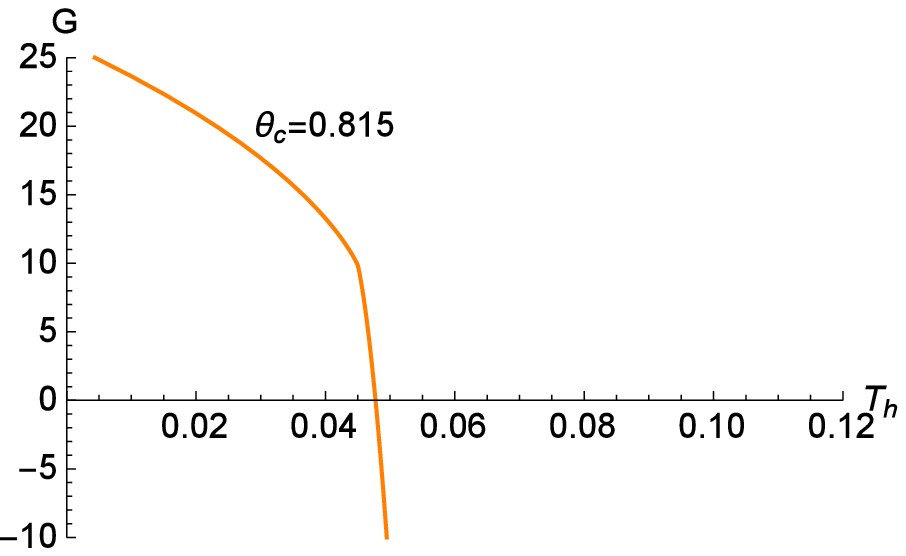} &
    \includegraphics[width=60mm]{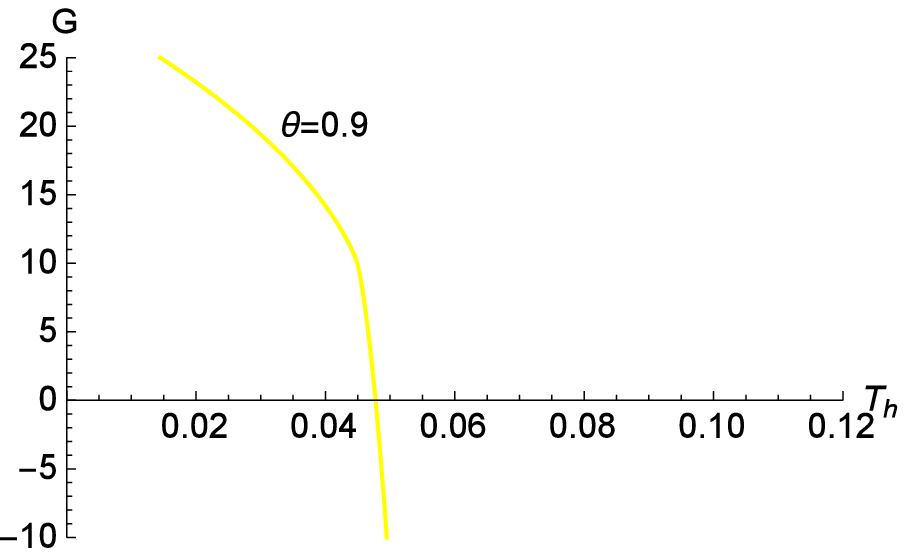} \\
    \includegraphics[width=60mm]{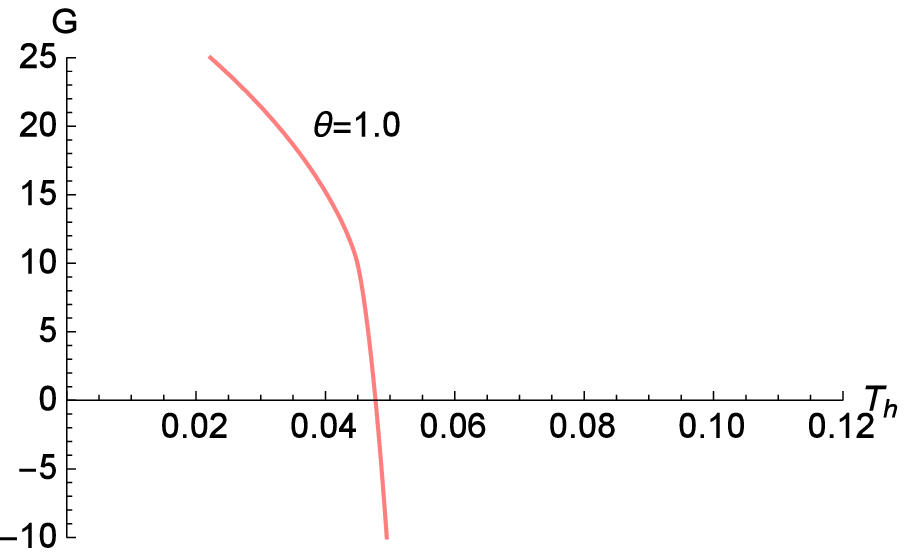} &
    \includegraphics[width=60mm]{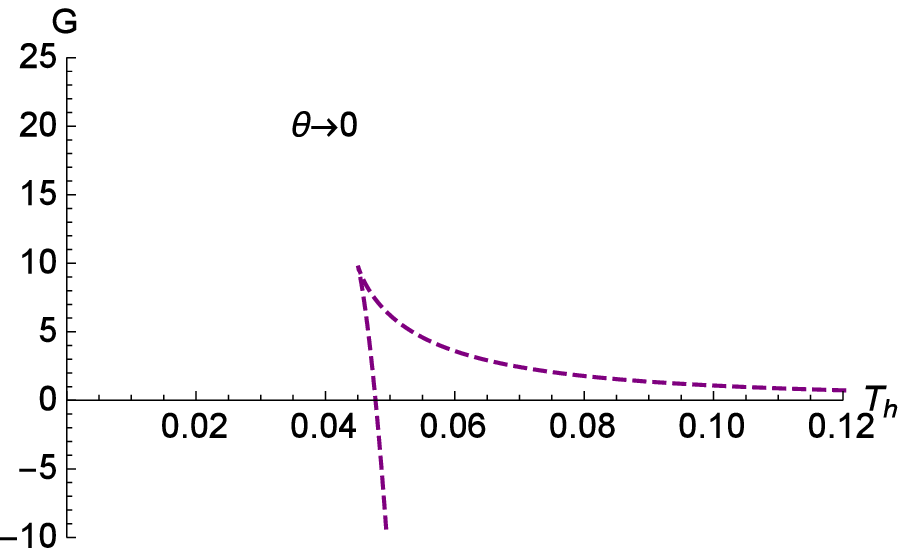}
  \end{tabular}
\end{center}
\caption{For $n=5$ and $k=3$, plots of the relation of $G$ with respect to $T_h$ at different noncommutative pressure $P_{NC}$ but at the constant pressure $P$ that corresponds to $l=10$. The dashed curve in the last diagram is the commutative case.}
\label{tu1}
\end{figure}

\begin{figure}
\centering
\includegraphics[width=70mm]{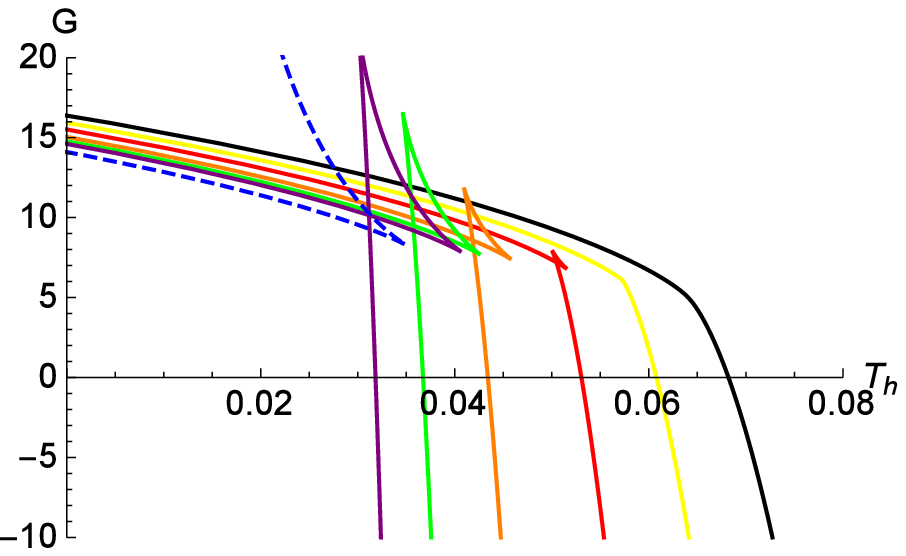}
\caption{For $n=5$ and $k=3$, plots of the relation of $G$ with respect to $T_h$ at the constant noncommutative pressure $P_{NC}$ that corresponds to $\theta=0.5$ but at different thermodynamic pressure $P$ that corresponds to $l=7$ $(\text{{\color{black}Black}})$, 7.855 $(\text{{\color{yellow}Yellow}})$, 9 $(\text{{\color{red}Red}})$, 11 $(\text{\textcolor[rgb]{1.00,0.50,0.00}{Orange}})$, 13 $(\text{{\color{green}Green}})$, and 15 $(\text{\textcolor[rgb]{0.50,0.00,0.51}{Purple}})$, respectively. The \text{{\color{blue}blue}} dashed curve is associated with the vanishing thermodynamic pressure that corresponds to the asymptotic Minkowski spacetime with $l\rightarrow\infty$.}
\label{tu2}
\end{figure}

The behaviors of the Gibbs free energy governed by eq.~(\ref{gibb}) are plotted in Figures \ref{tu1} and \ref{tu2}. In Figure \ref{tu1}, the phase transition is described for the constant thermodynamic pressure $P$ but the varying noncommutative pressure $P_{NC}$. The critical noncommutative pressure corresponds to $\theta_c=0.815$, see the fifth diagram of Figure \ref{tu1}. Once the noncommutative pressure $P_{NC}$ is lower than the critical noncommutative pressure, or the noncommutative parameter is larger than $\theta_c$, the characteristic swallowtail behavior disappears, which means that no first order phase transitions occur. In Figure \ref{tu2}, the phase transition is depicted for the constant noncommutative pressure $P_{NC}$ but the varying thermodynamic pressure $P$. The critical thermodynamic pressure corresponds to $l_c=7.855$, see the yellow curve in Figure \ref{tu2}. When the thermodynamic pressure $P$ is larger than the critical thermodynamic pressure, or the curvature radius of the AdS spacetime is smaller than $l_c=7.855$, no first order phase transitions occur. As a result, the thermodynamic pressure $P$ and noncommutative pressure $P_{NC}$  give the opposite contributions to the first order phase transition. Physically, $P$ dominates the infra-red regime (lower than its critical thermodynamic pressure) while $P_{NC}$ does the ultra-violet regime (larger than its critical noncommutative pressure). As our special examples, we also give the asymptotic behaviors of the Gibbs free energy under the limits $\theta\rightarrow 0$ and $l \rightarrow \infty$ in Figures \ref{tu1} and \ref{tu2}, respectively. The first limit leads to the case of the pure AdS spacetime without noncommutativity, and the second one to the noncommutative case without the AdS background as reported in ref.~\cite{AESS}.

From eq.~(\ref{gibb}) we notice a special point where the Gibbs free energy vanishes. This special point dubbed by $r_g$ satisfies the following equation,
\begin{equation}
r_g=\left[{\frac{1+G\left(n,k;\frac{r_g}{2\sqrt{\theta}}\right)}{1-G\left(n,k;\frac{r_g}{2\sqrt{\theta}}\right)}}\right]^{1/2}l.
\end{equation}
%where $x_g=r_g/(2\sqrt{\theta})$.
Under the commutative limit, it comes back to the known result $r_g=l$ given in ref.~\cite{BPD3}. At this point, the first order Hawking-Page phase transition shows up between the thermal radiation and the large black hole. If $r_h < r_g$, the noncommutative AdS spacetime is more stable, which means that the black hole will evaporate. If $r_h > r_g$, the Gibbs free energy is negative, indicating that the black hole is in a more stable thermodynamical configuration.

\section{Isoperimetric inequality}\label{sec3}
In the Euclidean space, the isoperimetric inequality implies that the geometric volume $\mathcal{V}$ and area $\mathcal{A}$ satisfy the following inequality for a given $(n-1)$-dimensional connected domain,
\begin{equation}
\mathcal{R}:=\left[\frac{(n-1)\mathcal{V}}{\omega}\right]^{\frac{1}{n-1}}\cdot \left(\frac{\omega}{\mathcal{A}}\right)^{\frac{1}{n-2}}\leq 1,
\end{equation}
where the equality holds if and only if the domain is a standard round ball.

As a counterpart, the reverse isoperimetric inequality has been asserted~\cite{CGKP,AKMS} in any asymptotic AdS black holes,
\begin{equation}
\mathcal{R}:=\left[\frac{(n-1)V}{\omega}\right]^{\frac{1}{n-1}}\cdot \left(\frac{\omega}{A}\right)^{\frac{1}{n-2}}\geq 1, \label{reiso}
\end{equation}
where $V$ denotes the thermodynamic volume and $A$  the horizon area of black holes.
Note that the equality is attained for the ordinary Schwarzschild-AdS black hole.

Physically, the reverse isoperimetric inequality indicates that the entropy of black holes is maximized for the Schwarzschild-AdS spacetime at a given thermodynamic volume.

Up to now, the above statement has been verified for a variety of black holes with the horizon of spherical topology~\cite{CGKP} and black rings with the horizon of toroidal topology~\cite{AKMS}. The only counterexample, as we know, is the ultra-spinning black hole~\cite{hm}. However, the reverse isoperimetric inequality still remains valid for noncommutative black holes with smeared matter distributions. Let us check it.

In our case, with the help of the entropy (eq.~(\ref{entro})) and the thermodynamic volume (eq.~(\ref{volu})) of noncommutative black holes, we can obtain
\begin{equation}
\mathcal{R}=\left[\frac{\Gamma\left(\frac{n+k-1}{2}\right)}{\gamma\left(\frac{n+k-1}{2},\frac{r_h^2}{4\theta}\right)}\right]^{\frac{1}{(n-1)(n-2)}}.
\end{equation}
Considering the property of the lower incomplete gamma function, we yield $\mathcal{R}\geq 1$ and the equality is attained for taking the commutative limit $\theta\rightarrow 0$, i.e. for the ordinary Schwarzschild-AdS black hole. In other words, for a black hole with a kind of  smeared matter distributions, its entropy is smaller than that of the black hole without any distributions at the same thermodynamic volume.

Next, we turn to the discussion about which kind of distributions corresponds to the maximum entropy for the black holes with smeared matter distributions. We compare the two kinds of distributions with the power index $k$ and $k+1$, respectively, see eq.~(\ref{fenbu}), in the $n$-dimensional spacetime. The ratios are denoted by $\mathcal{R}_{n,k}$ and $\mathcal{R}_{n,k+1}$, respectively, and the ratio of the two ratios can be calculated to be
\begin{equation}
\frac{\mathcal{R}_{n,k}}{\mathcal{R}_{n,k+1}}=\left[\frac{\Gamma\left(\frac{n+k-1}{2}\right)}{\Gamma\left(\frac{n+k}{2}\right)}\cdot
\frac{\gamma\left(\frac{n+k}{2},\frac{r_h^2}{4\theta}\right)}{\gamma\left(\frac{n+k-1}{2},\frac{r_h^2}{4\theta}\right)}
\right]^{\frac{1}{(n-1)(n-2)}}.
\end{equation}
We can prove $0< \mathcal{R}_{n,k}/\mathcal{R}_{n,k+1}< 1$, which means that the black hole with the Gaussian smeared matter  distribution ($k=0$) holds the maximum entropy for a fixed spacetime dimension at a given thermodynamic volume, as expected. In other words, the more concentrated the matter distribution of black holes is, the greater the entropy of black holes is.

\section{Conclusion}\label{sec4}
Based on our recent work~\cite{MX} about the high-dimensional Schwarzschild-Tangherlini AdS black hole with the non-Gaussian smeared matter distribution, we deal with the noncommutative parameter as an independent thermodynamic variable called the  noncommutative pressure. As a result, the noncommutative pressure together with its conjugate volume appears in the generalized Smarr relation and its variation comes out in the first law of thermodynamics. Through analyzing the Gibbs free energy, we point out that the noncommutative pressure and the thermodynamic pressure make the opposite effects in the phase transition of noncommutative black holes. Physically, the noncommutative pressure dominates the UV physics while the thermodynamic pressure does the IR physics. We also discuss the first order Hawking-Page phase transition. Furthermore, we calculate the reverse isoperimetric inequality for the noncommutative black holes and indicate that the noncommutative black hole with the Gaussian smeared matter distribution ($k=0$) holds the maximum entropy at a fixed thermodynamic volume.

\section*{Acknowledgments}
Z.-M. Xu would like to thank Y.-M. Wu and L. Zhao for their helpful discussions.
This work was supported in part by the National Natural
Science Foundation of China under grant No.11675081. At last, the authors would like to thank the anonymous referee for the helpful comment that indeed greatly improves this work.

\end{document}